\documentclass[]{spie}  

\usepackage{amsmath,amsfonts,amssymb}
\usepackage{graphicx}
\usepackage[colorlinks=true, allcolors=blue]{hyperref}
\usepackage{adjustbox}
\usepackage{indentfirst}

\usepackage{xcolor}

\usepackage{etoolbox}
\apptocmd{\thebibliography}{\setlength{\itemsep}{10pt}}{}{}

\title{Bronchoscopic video synchronization for interactive multimodal inspection of bronchial lesions}

\author{Qi Chang,$^a$ Patrick D. Byrnes,$^a$ Danish Ahmad,$^b$ Jennifer Toth,$^b$\\
\vspace*{-10pt}Rebecca Bascom $^b$ and William E. Higgins $^{a,*}$\\
\vspace*{8pt}$^a$School of Electrical Engineering and Computer Science\\
$^b$College of Medicine\\
Penn State University, University Park and Hershey, PA
}

\authorinfo{$^*$Correspondence: Email: weh2@psu.edu;
	WWW: http://mipl.ee.psu.edu/; Telephone: 814-865-0186; Fax: 814-863-5341.}

\begin{document} 
\maketitle

\begin{abstract}
\noindent With lung cancer being the most fatal cancer worldwide, it is important to detect the disease early.
A potentially effective way of detecting early cancer lesions developing along the airway walls (epithelium) is bronchoscopy.  
To this end, developments in bronchoscopy offer three promising noninvasive modalities for imaging bronchial lesions: white-light bronchoscopy (WLB), autofluorescence bronchoscopy (AFB), and narrow-band imaging (NBI).
While these modalities give complementary views of the airway epithelium, the physician must manually inspect each video stream produced by a given modality to locate the suspect cancer lesions.
Unfortunately, no effort has been made to rectify this situation by providing efficient quantitative and visual tools for analyzing these video streams. 
This makes the lesion search process extremely time-consuming and error-prone, thereby making it impractical to utilize these rich data sources effectively.
We propose a framework for synchronizing multiple bronchoscopic videos to enable interactive multimodal analysis of bronchial lesions.
Our methods first register the video streams to a reference 3D chest computed-tomography (CT) scan to produce multimodal linkages to the airway tree.
Our methods then temporally correlate the videos to one another to enable synchronous visualization of the resulting multimodal data set.
Pictorial and quantitative results illustrate the potential of the methods.

\end{abstract}


\section{INTRODUCTION}
\label{sec:intro}  

Given that lung cancer is the most common cause of cancer death in the world, an important goal in managing lung cancer is to detect the disease early.
To this point, many early cancerous lesions develop along the airway walls (epithelium). 
A potentially effective way for detecting such lesions is to use bronchoscopy to noninvasively image epithelial lesions \cite{Wisnivesky-Chest2013}.
Three complementary modalities have recently been developed for this purpose \cite{Inage2018early,myers2017early}:
white-light bronchoscopy (WLB), autofluorescence bronchoscopy (AFB), and narrow band imaging (NBI). 
WLB, the standard video mode for general bronchoscopic procedures, considers the reflection of broadband visible light to image bronchial surfaces.
AFB highlights autofluorescence differences between the normal epithelium and developing epithelial lesions.
NBI uses narrow band filtered light to enable focused visualization of the imaged airway wall's vessel structure.
In principle, the physician could perform a multimodal bronchoscopic airway exam on a patient to search for bronchial lesions by inspecting the endoluminal video stream collected by each modality.
Preliminary research indicates that this collection of videos gives a rich multimodal data source enabling more robust lesion detection than any single mode separately  \cite{Inage2018early,myers2017early}.

Unfortunately, the physician must perform the airway exam for each modality separately, thereby giving several disjoint data sources not directly linked.
Furthermore, the physician must manually inspect each data source one at a time and interactively define lesion confirmations at airway sites of interest.
At present, this task is not only impractical but also error prone and likely to result in missed lesion sites \cite{Dumas-TARD2016,myers2017early}.
As a step toward addressing these problems, the video streams can be recorded for later off-line review, but there still is no available means for relating individual video sources and linking multimodal video findings at correlated sites.
Thus, methods still do not exist to help exploit the considerable potential of multimodal bronchoscopy for lesion detection.

{\it To address this need,  we propose a framework for synchronizing multiple bronchoscopic videos to enable an interactive multimodal analysis of bronchial lesions.}
The method draws upon multimodal CT-video registration to make a preliminary registration of each video source to a reference patient 3D chest CT scan.
Next, a K-D tree search synchronizes the registered videos to enable multimodal playback.
Section \ref{sec:methods} describes our methodology. Section \ref{sec:results} next illustrates results, while Section \ref{sec:discussion} gives final comments.

\section{METHODS}
\label{sec:methods}  
The basic input is a series of four bronchoscopic videos collected during a multimodal airway exam of the two lungs.
These exam videos were collected with two separate bronchoscopes.
The first bronchoscope provides high-definition
WLB video and related NBI video.
The second device accommodates dual-mode autofluorescence bronchoscopy.
Two airway exams are also completed with this device to give separate AFB and WLB videos.
These devices enable a detailed search for suspect early cancer lesions arising along the airway epithelium\cite{Inage2018early,myers2017early}.

We note that the videos produced by the physician's major airway exam  of the two lungs offer considerable detail on the airway endoluminal structure, as later examples show.
Nevertheless, the process of synchronizing all four videos simultaneously in time, such that the videos can be played to show the same airway locations together, is very difficult.
Three reasons explain this circumstance.  
First, airway exam videos, while approximately observing the same airways, can vary in time.  
Second, even if the videos have a similar time frame, they most definitely {\it never observe the exact same location with exactly the same field of view.} 
Finally, the endpoints of the terminal airways, before the physician backs up to observe a different lung segment, can differ.

We propose two distinct approaches for partially rectifying these issues.
A basic synchronization approach entails interactive synchronization of each video with a reference CT-based virtual bronchoscopic (VB) movie sequence through the airways at selected temporal segments. 
A more advanced synchronization approach draws upon automatic video parsing, semi-automatic key-frame registration, automatic CT-video sequence registration, and a final automatic search to enable more precise synchronization.
We describe these methods below. 

\subsection{Overview of the Process}
\label{subsec:methods_overview}
Our methods take a patient's 3D chest computed tomography (CT) scan and a series of prerecorded multimodal airway exam videos as inputs.
The basic process followed by our methods is the following:
\vspace*{-6pt}
\begin{enumerate}
\itemsep=-3pt
\item  Using the patient's chest CT scan, a chest model, consisting of the airway tree, airway endoluminal surfaces, and airway centerlines, is constructed automatically.
\item  The physician performs a multimodal airway exam on the patient.  
In particular, the physician first uses a dual-mode WLB-NBI bronchoscope, performing one complete central airway exam for each modality.
Next, the physician employs a dual-mode WLB-AFB bronchoscope to perform another similar dual-mode set of airway exams.
The video streams are recorded for subsequent use.
\item  Using the chest CT information and recorded video streams, a semi-automatic process synchronizes the video sources to the chest CT model for a selected airway segment of interest.
\item The physician interacts with the synchronized multimodal data set using a custom interactive system.
\end{enumerate}
\vspace*{-6pt}

\begin{figure} [!htb]
	\vspace*{-2pt}
\begin{center}
   \begin{tabular}{ccc}
		\includegraphics[width=1.8in]{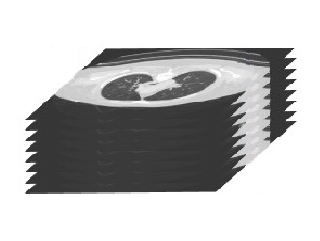}&
		\includegraphics[width=2.0in]{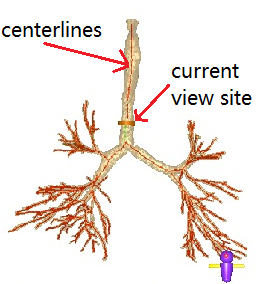}&
		\includegraphics[width=1.6in]{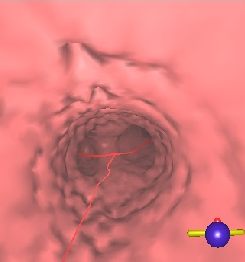}\\
(a) 3D CT Scan &  (b) CT-based airway tree & (c) VB view \\
    \end{tabular}
	\end{center}
	\vspace*{-4pt}
	\caption{\small \baselineskip=11pt Construction of the chest model: (a) 3D CT scan, (b) CT-based airway tree and centerlines, (c) airway endoluminal surfaces (VB view). For each patient, we use their 3D CT scan to construct a virtual airway tree and then calculate centerlines shown as red lines inside the pink CT-based segmented airway tree. In addition, the yellow ``disk" icon in the 3D airway tree view denotes the current viewing position.}
	\label{fig:CTDemo}
	\vspace*{-2pt}
\end{figure}

Regarding step 1, we use previously validated methods to construct the chest model as highlighted in \cite{Byrnes-IEEEBME2019}.
The airway-tree centerline structure spans the airways captured within the CT-based segmented airway tree, as shown in Fig. \ref{fig:CTDemo}.
This structure is made up of interconnected centerlines that define paths traversing the airways.
The centerlines in turn are made up of a discrete set of view sites.   
Our goal is to correlate the various videos to the CT-based airway tree along a pre-selected airway segment within the tree, which enables the user to check the same view sites from different video streams.
\\

Regarding the airway exam of step 2, the physician produces a pair of separate video streams for WLB and NBI using the first bronchoscope and a second pair of distinct video streams for WLB and AFB using the second bronchoscope.
Each video constituting a given pair has the same device field of view --- but the videos are most definitely not aligned in anyway!
Subsection \ref{subsec:methods_basis}-\ref{subsec:methods_synchronization} addresses steps 3 and 4. In particular, Subsection \ref{subsec:methods_basis} introduces basic system-level concepts. Section \ref{sec:results} gives examples of multimodal video interaction.

\subsection{System Basis}
\label{subsec:methods_basis}
We first note that, while each airway exam video approximately covers the same central airway structure, their actual constituent video frames can vary considerably and be different for every video frame in a sequence in terms of the observed field of view, bronchoscope motion, and beginning and ending airway tree branch points.
Hence, the multi-video synchronization problem has many challenges.
To address these challenges, we use the CT chest model as the reference standard for linking all videos together.
More specifically, the 3D CT-based chest model acts as a virtual space that is directly equivalent to the real patient/bronchoscope chest space\cite{Merritt2013}.  
In particular, we directly correlate CT-based virtual bronchoscope (VB) views (also known as endoluminal renderings) to the real bronchoscopic video frames imaging the (interior) endoluminal airway anatomy.
This idea is the fundamental principle driving recently introduced image-guided bronchoscopy systems \cite{Reynisson2014}.

Thus, our synchronization approaches use the image-guided bronchoscopy concept of synchronizing the "real" video stream data to corresponding locations within the CT-based virtual airway anatomy.
We do this by using one of two main ideas:  (1) linking a video frame through the CT-based centerlines;
(2) linking a video frame through direct CT-video registration of a video frame to a VB view.
Therefore, given the CT-based chest model and 4 input video streams, our methods proceed with the processing discussed below.

\subsection{Video Synchronization}
\label{subsec:methods_synchronization}
\subsubsection{Basic Method}
To begin basic synchronization, we first preprocess the given four video streams via the following steps: (1) interactively segment each video stream into subsequences according to time stamps and (2) associate the start and end frames of each subsequence with the CT model. 
To select the time stamps, the user plays a given video and selects key locations, such as the main carina or other landmark bifurcation views.
Then, the user navigates a VB viewer near each position for these video time stamps and semi-automatically registers a VB view to a video frame. 

The user next selects an airway path of interest.
The path corresponds to a portion of the airway exam performed by the physician.
Denote this path as \ ${\bf p} = \left\{{\bf v}_1, {\bf v}_2, {\bf v}_3, \ldots, {\bf v}_{N-1}, \, {\bf v}_N \right\}$ . 
Each ${\bf v}_i, i = 1, \ldots, N,$ is a view site defined by: a) a 6-parameter vector specifying the 3D $(x,y,z)$ position and 3
orientation angles $(\alpha, \beta, \gamma)$; and b) an up vector, which defines the orientation of the imaging camera (the virtual or real bronchoscope).
The goal now is to correlate the available subsequences for each video to this path to give a synchronized multimodal data structure, whereby all start and end points of subsequences are registered to the corresponding CT-based position.
The details of the preprocessing are shown below:

\vspace*{3pt}
\noindent \underline{A. Preprocessing:}\\
\noindent  For each video $I_j,\ j = 1, 2, 3, 4$\\
\hspace*{0.15in} Interactively pick a small set of time stamps \, $\{ t_a, t_b, t_c, \ldots \}$ \,
   spanning the airway path \, ${\bf p}$\\
\hspace*{0.15in} Load the video into Adobe Premiere and make a cut at each time stamp \\
\hspace*{0.15in} For each consecutive pair of time stamps $(t_i, \, t_{i+1}) \, \in \, \{ t_a, t_b, t_c, \ldots \}$\\
\hspace*{0.3in}  Semi-automatically register the CT virtual bronchoscope to view sites
   $({\bf v}_i, \, {\bf v}_{i+1})$ near each time stamp\\
\hspace*{0.3in}  Associate $I_j$'s video subsequence between $\{ t_i, \, t_{i+1}\}$
   to the subset of view sites $({\bf v}_i, \ldots {\bf v}_{i+1})$

\vspace*{3pt}

So far, we have preprocessed all data to form rough subsequence linkages between the CT model and 4 distinct video sources.
In general, the set of time stamps is sufficient to span the path of interest.
We next select a path subsegment \, $\left\{ {\bf v}_j, {\bf v}_{j+1}, \ldots, {\bf v}_k \right\} \subset {\bf p}$ \,
for further processing below:

\vspace*{3pt}

\label{alg:basic}
\noindent \underline{B. Basic synchronization:}\\
\noindent    Select a CT-based virtual bronchoscope sequence between $({\bf v}_i, \ldots {\bf v}_{i+1})$\\
\noindent  For a designated reference video $I_1$ \ \verb+[e.g., use the WLB video]+\\
\hspace*{0.15in} For each designated subsequence between $\{ t_i, \, t_{i+1}\}$\\
\hspace*{0.3in}  Assign the subsequence in Adobe Premiere according to the order of $({\bf v}_i, \ldots {\bf v}_{i+1})$\\
\noindent For each remaining video $I_j,\ j = 2, 3, 4$ \\
\hspace*{0.15in} For the subsequence between $\{ t_i, \, t_{i+1}\}$ according to the order of $({\bf v}_i, \ldots {\bf v}_{i+1})$\\
\hspace*{0.3in}  Interactively modify the video play speed and cut redundant actions between
   $\{ t_i, \, t_{i+1}\}$\\
\hspace*{0.4in}  (so that $t_i$ and $t_{i+1}$ can maintain consistency with time stamps in the reference video)\\
\hspace*{0.3in}  Assign the subsequence in the associate video track of Adobe Premiere

\vspace*{3pt}

The above method provides a complete, albeit crude, synchronization of all videos to the reference CT model.
In particular, the only registered frames are those corresponding to selected time stamps;
frames within subsequences are not registered and may not correspond smoothly to the CT model or to the other video sources.
Also, small frame shots within subsequences might entail motion unlike the motion of other constituent subsequence video frames;
this degrades the synchronization locally.
Finally, the multimodal data set can only be played in the forward direction, as suitable smooth linkages for reverse play are not available.

\subsubsection{Advanced Method}
To improve upon the basic method, we propose an advanced synchronization method that registers the video frames of each video stream to the CT model using the multi-frame registration method of Byrnes and Higgins \cite{Byrnes-ISBI2020}. 
The overview of this method is described below:
\begin{enumerate}
\itemsep=-3pt
\item Binary Robust Independent Elementary Features (BRIEF) are extracted for all frames and then are used to analyze the endobronchial video shots.
\item The BRIEF features and shots are used to construct a geometric model describing incremental bronchoscope motion.
\item Given the information above, a sparse set of key frames are parsed to represent the endobronchial video shots.
\item The user interactively applies automatic registration to a subset of key frames, chosen either at the start or the end key frame of the video shots, to the CT-based airway tree.
This initializes the registration process.
\item Drawing on the matched features and analyzed video motion, the frames within each video shot are registered to the 3D airway tree \cite{Byrnes-ISBI2020}.
\item Interactively check the mappings for registered frames and modify key frames until there are no obvious misalignments.
In addition, the user can add key frames into the shots and use them as the registration reference. 
\end{enumerate}

Steps 1-3 are performed using the video parsing approach of Byrnes and Higgins\cite{Byrnes-IEEEBME2019}. 
Step 4 is accomplished by Merritt {\it et al} \cite{Merritt2013}.
Since there can be some uninformative frames that have little use, the user can manually delete them to guarantee the registration process\cite{Byrnes-IEEEBME2019}.
Further improvements can be made by detecting uninformative or excessive redundant video frames according to bronchoscope modality\cite{chang2020autofluorescence,mctaggart2019robust}.

Given these registrations, we can now instead associate individual video frames from the various video sources to the virtual bronchoscope view sites along the CT-based airway tree.
This dispenses with the need for dealing with the varying duration of video subsequences.
Furthermore, we modify the video format without changing any frame information for the convenience of loading any frames from the video streams in real-time.
Since the popular image processing tool "OpenCV" is not suitable for our needs, the more basic library "FFmpeg" was used for both converting video format and loading frame information\cite{tomar2006converting}.
Based on these well-registered frames, we propose advanced synchronization as follows.

\vspace*{10pt}

\label{alg:advanced}
\noindent 2. \underline{Advanced synchronization:}\\
\noindent For each video $I_j,\ j = 1, 2, 3, 4$ \\
\hspace*{0.15in} Modify the video format \verb+[e.g., GOP size 12, I/P frames only]+\\
\noindent For reference video $I_1$ \ \verb+[e.g., use the WLB video of scope]+\\
\hspace*{0.15in} Create a target list of frame numbers from the beginning to the end of the reference video\\
\hspace*{0.15in} Store the  frame numbers of well-registered frames and their pose information as a reference in a K-D tree\\
\hspace*{0.15in} For each remaining video $I_j,\ j = 2, 3, 4$\\
\hspace*{0.3in} Search the nearest view site by using K-D tree and store the associated frame numbers on a target list\\
\hspace*{0.3in} Use an independent thread to search and store frames in a limited size container\\
\noindent Use any display button from the window of the reference video to (forward/backward) display 

\vspace*{10pt}

Specifically, a large size of Group Of Picture (GOP) and mixture with randomly occurred bi-directional predicted frames (B frames) make the backward displaying of a video source in real-time become very difficult, based on the MPEG decoding mechanism.
Here, we modify the GOP size of 12 frames with only intra-coded frames (I frames) and predicted frames (P frames).
In addition, a limited size container that can store the information for 10-frame blocks was set for each video stream, so as to both smooth the switchover and control the frame rate of the displayed video.
Thus, this method now permits both forward and backward play through the multimodal data set. 

\section{RESULTS}
\label{sec:results}  
\subsection{Basic synchronization}
\label{subsec:results_basic}
\begin{figure} [!htb]
	\vspace*{-2pt}
\begin{center}
    
    \hspace*{-0.1in}\begin{tabular}{c}
		\includegraphics[width=\textwidth]{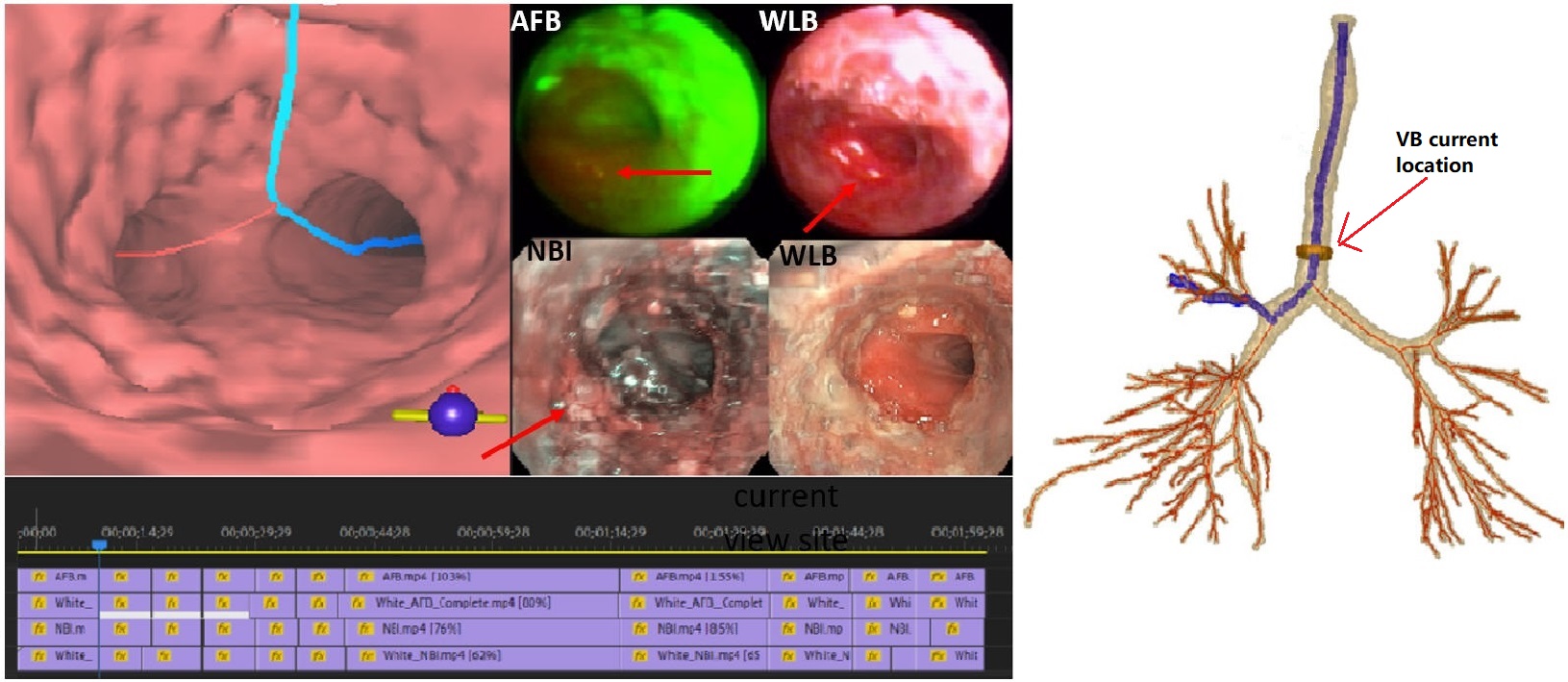} \\
	(a) System view at a synchronized time stamp. \vspace*{6pt} \\
		\includegraphics[width=\textwidth]{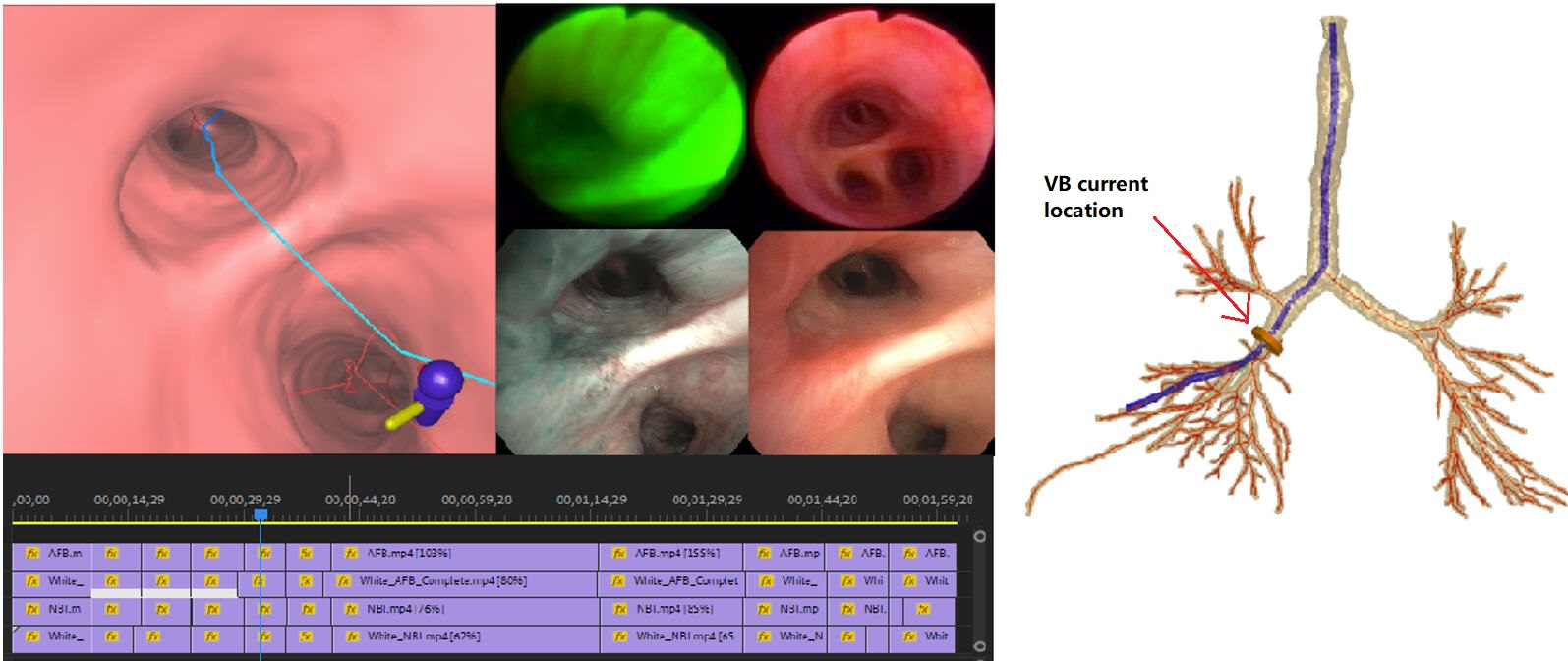} \\
    (b) System view at a site between time stamps. \\
    \end{tabular}
\end{center}
	\vspace*{-4pt}
	\caption{\small \baselineskip=11pt Basic multi-video viewing system using case 21405-171:  (a) at a time stamp of a key site; (b) at a site between time stamps. The left panel shows five endoluminal views at approximately the same airway location with time stamps and video sub-sequences on the bottom. The purple blocks depict the video profiles interactively segmented into subsequences for each video track and made consistent with the VB sequence. In addition, the blue cursor in the video-sequence panel indicates the current displayed time stamp. The right panel depicts the reference 3D CT-based airway tree view with highlighted blue path of interest and a red arrow that points to the current VB location. WLB bronchoscopic view from the AFB scope is omitted since it has a similar view to WLB from the NBI scope.}
	\label{fig:Basic}
	\vspace*{-2pt}
\end{figure}

Using patient data from lung-cancer patient case 21405-171, screenshots are shown to illustrate the algorithm for basic synchronization.
For each example in Fig. \ref{fig:Basic}, the composite views depict the following.
The left panel shows five endoluminal views at approximately the same airway location: a large CT-based endoluminal VB view and four distinct video frames [top pair --- WLB-AFB scope; bottom pair --- WLB-NBI scope]; 
bottom portion of panel plots each video sequence broken up into separate sub-sequences shown in purple blocks, with the blue cursor indicating the current play time.
The right panel depicts the reference 3D CT-based airway tree view.  The blue line in all views signifies the path of interest, where Fig. \ref{fig:Basic}a shows a view site at the main carina and Fig. \ref{fig:Basic}b is at a location in between two registered time stamps.
All views are relative to this viewing position. 

Fig. \ref{fig:Basic}a shows that all views are well synchronized at a time stamp location.
Fig. \ref{fig:Basic}b, however, shows a similar basic view for a view site interior to a video subsequence.
Clearly, all video views don't appear especially well aligned to the precise same view site.
Nevertheless, this basic approach does give useful multimodal video alignment information.

\subsection{Advanced synchronization}
\label{subsec:results_advan}
In this subsection, we first use cancer patient data from case 21405-171 to make a comparison to the basic synchronization method.
Then, the results of the advanced synchronization method from case 21405-173 illustrate the potential of the method.

\subsubsection{Case 21405-171}
\begin{figure} [!htb]
		\hspace{-0.1in}\begin{tabular}{c}
        \includegraphics[width=\textwidth]{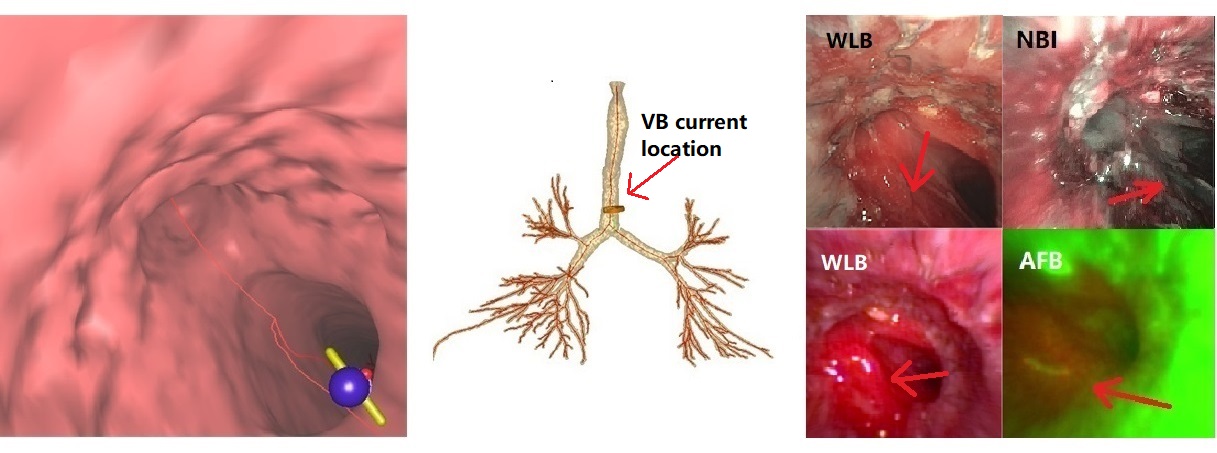} \\
		(a) Same view site as shown in Fig. \ref{fig:Basic}a.\\
		\includegraphics[width=\textwidth]{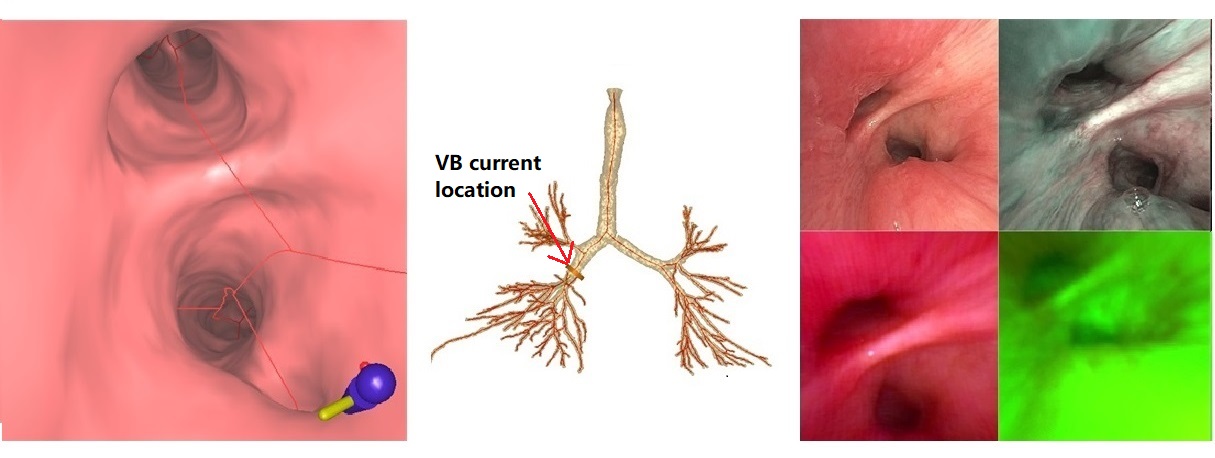}\\
        (b) Same view site as shown in Fig. \ref{fig:Basic}b.
        	
		\end{tabular}

	\caption[]{\baselineskip=12pt Synchronized views using the advanced method for case 21405-171. (a) Same view site as shown in Fig. \ref{fig:Basic}a. (b) Same view site as shown in Fig. \ref{fig:Basic}b. From left to right is: CT-based VB view, CT-based airway tree and 4 distinct video frames. The red arrow in the middle of each sub-figure points to the current VB location. Obviously, Fig. \ref{fig:Advanced}b shows a better view comparing to Fig. \ref{fig:Basic}b.}
	\label{fig:Advanced}
	\vspace*{-1pt}
\end{figure}

For each example in Fig. \ref{fig:Advanced}, the composite views depict the following.
The left panel shows a reference CT-based endoluminal VB view.
The middle panel depicts the reference 3D CT-based airway tree and indicates the VB's current position.
The right panel presents four endoluminal views at approximately the same airway location.
[top pair -- WLB-NBI scope; bottom pair -- WLB-AFB pair].
For each frame, the view is cropped after correcting for bronchoscope distortion.
These views correspond to the same view site as considered in Fig. \ref{fig:Basic}, which used the basic method.

\vspace*{-2pt}

Proper alignment of all frames with the reference CT-based VB views is now apparent. 
In addition, the advanced method facilitates full access to all frames.
For example, the bottom-left WLB frame depicted in Fig. \ref{fig:Advanced}b cannot be found in the subsequence defined via the basic synchronization method.
Yet, given the registrations performed for the advanced method, this frame is now easily located.

\subsubsection{Case 21405-173}
\begin{figure} [!htb]
		\hspace*{16pt}
		\hspace{-0.1in}\begin{tabular}{c}

        \includegraphics[width=0.96\textwidth]{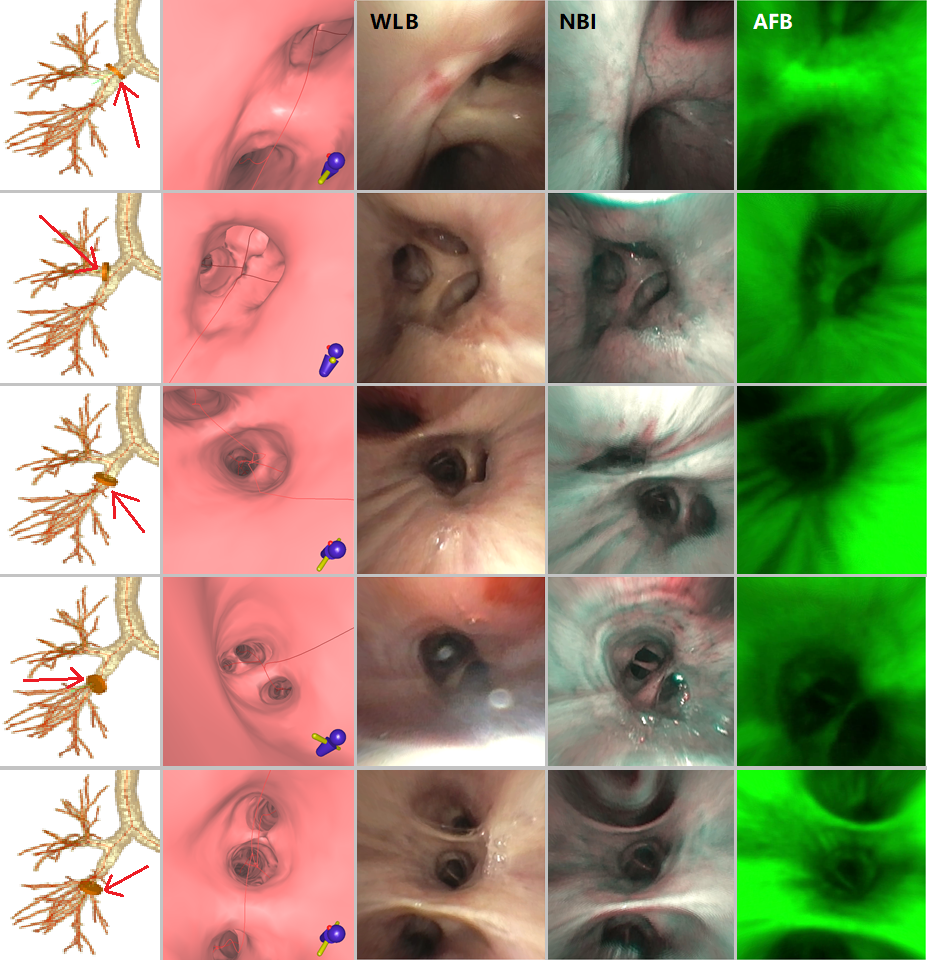} 
        	
		\end{tabular}

	\caption[]{\baselineskip=12pt Synchronized views using the advanced method for case 21405-173.
	Each row, from top to bottom, gives different view sites: right main bronchi, right upper lobe bronchus, right intermediate bronchus, right middle lobe bronchus and right lower lobe bronchus.
	Each column, from left to right, depicts the same view site in the 3D airway tree, VB view, bronchoscopic views of WLB-NBI scope pair and the AFB bronchoscopic view. WLB view from the AFB scope is omitted for clarity. The red arrows in the first column show the current VB locations.}
	\label{fig:Demo}
	\vspace*{-1pt}
\end{figure}

Fig. \ref{fig:Demo} presents composite views using the advanced synchronization system for case 21405-173. In particular, only an airway segment within the right lung of the airway exam videos is used.
For each row from top to bottom, the figure shows different view sites: right main bronchi, right upper lobe bronchus, right intermediate bronchus, right middle lobe bronchus and right lower lobe bronchus, to fully illustrate the synchronized video sequences.
In each row, the figure gives the view site in 3D airway tree, VB view, as well as corrected frames from bronchoscopic views of WLB-NBI scope pair and the AFB bronchoscopic view, from left to right. 

As is apparent in Fig. \ref{fig:Demo}, all views are well synchronized at the various airway tree sites.
More specifically, we used the WLB source from the NBI scope as our reference video with 1185 well-registered frames.
Among these 1185 frames, 1114 frames, 1003 frames and 1000 frames are found to be associated, respectively, to corresponding frames in NBI video and bronchoscopic videos (WLB, AFB) from the AFB scope. These results for registration processing are shown in Table. \ref{table:register}.

\begin{table}[ht]
\vspace*{1pt}\resizebox{\textwidth}{!}{\begin{tabular}{|c|c|c|c|c|c|}
\hline
\multicolumn{1}{|l|}{\textbf{Video Type}} & \multicolumn{1}{l|}{\textbf{Total Frames}} & \multicolumn{1}{l|}{\textbf{Keyframes}} & \multicolumn{1}{l|}{\textbf{Interactively Registered}} & \multicolumn{1}{l|}{\textbf{Sequence Registered}} & \multicolumn{1}{l|}{\textbf{Associated}} \\ \hline
\textbf{NBI-WLB}                          & 1528                                       & 154                                     & 18                                                     & 1185                                              & N/A                                      \\ \hline
\textbf{NBI}                              & 1408                                       & 161                                     & 17                                                     & 1100                                              & 1114                                     \\ \hline
\textbf{AFB-WLB}                          & 1528                                       & 138                                     & 22                                                     & 1000                                              & 1003                                     \\ \hline
\textbf{AFB}                              & 1618                                       & 86                                      & 18                                                     & 725                                               & 1000                                     \\ \hline
\end{tabular}}
\hspace*{3pt}
\caption[]{\baselineskip=12pt Synchronized view data using the advanced method on case 21405-173.
"NBI-WLB" means that we use the WLB source from the NBI scope as the reference standard in this example.
"Total" = total number of frames in the video, 
"Keyframes" = key frames parsed by Byrnes and Higgins's method \cite{Byrnes-ISBI2020},
"Interactively Registered" = key frames selected to be registered to the CT model,
"Sequence Registered" = sequence frames registered to the CT model based on these key frames\cite{Byrnes-ISBI2020},
"Associated" = video frames well-synchronized to the reference standard (it allows the repeated use from sequence registered frames).
}
\label{table:register}
\end{table}

Yet, user interaction time is still an issue.
In particular, when performing the advanced synchronization method, we point out that the actual time to complete a study is significant.
While parsing the videos to generate key frames is automatic, the step of registering the selected key frames to the CT chest model takes on the order of 1.5 hours for approximately 20 key frames.
Continuing, CT-video sequence registration is again automatic, taking roughly 2 minutes for over 1,000 frames.
Yet, the final interactive step 6 requires approximate 10 minutes for each video sequence.
All told, the time to fully process one video sequence takes on the order of 2.5 hours. 

Overall, the advanced synchronization method gives complete access to all frames during the airway exams, regardless of the mismatches during the time stamps provided by the basic synchronization method.
These pictorial and quantitative results illustrate the potential of using the advanced synchronization method as a tool to detect lesions.

\section{DISCUSSION}
\label{sec:discussion}  
Multimodal bronchoscopy offers a highly promising means for detecting early cancer lesions arising along the airway walls.
Yet, no tools currently exist to make this cancer detection method practical.
Our methodology provides a potentially useful and practical mechanism for detecting and processing bronchial lesions forming along the airway walls.
For the basic synchronization method, the user can easily operate in Adobe Premiere and have a general overview of lesion location. 
In order to make a specific and accurate diagnosis of the lesions and detailed detection for lesions through the airways, the advanced synchronization method has better potential. 
Future algorithmic work will focus on two fronts.
First, we must create a more automatic link between the parsed video sequences and the subsequent CT-video 
sequence registration step.
Second, CT-video sequence registration needs to be 
made more robust.
The major goal here is to further reduce the interaction
required to synchronize the videos.
Finally, other work will focus on using the synchronized video mechanism for automated lesion detection.

\acknowledgments 
This work was funded by NIH NCI grant R01-CA151433.
Dr. Higgins and Penn State have financial interests in Broncus Medical, Inc. These financial interests have been reviewed by the University’s Institutional and Individual Conflict of Interest Committees and are currently being managed by the University and reported to the NIH.

\newpage

\bibliography{bibtex/higgins,bibtex/higginsnew2019,bibtex/mipl} 
\bibliographystyle{spiebib} 


\end{document}